\documentclass[aps,a4paper,english,superscriptaddress,nofootinbib]{revtex4}

\usepackage{amsmath,amssymb,graphicx}

\makeatletter
\usepackage{babel}
\usepackage[active]{srcltx}
\usepackage{multirow}
\usepackage{graphicx,color}
\usepackage{changebar}
\usepackage{hyperref}
\usepackage[utf8]{inputenc}
\usepackage[T1]{fontenc}
\usepackage{ae}
\usepackage[caption = false]{subfig}

\usepackage{graphicx}
\usepackage{amsfonts}
\newcommand{\e}{\text{e}}

\newcommand{\be}{\begin{equation}}
	\newcommand{\ee}{\end{equation}}
\newcommand{\bea}{\begin{eqnarray}}
	\newcommand{\eea}{\end{eqnarray}}

\begin{document}

\title{Quantum information of the Aharanov-Bohm ring with Yukawa interaction in the presence of disclination}

\author{C. O. Edet}
\email[]{E-mail: collinsokonedet@gmail.com}
\affiliation{Faculty of Electronic Engineering Technology, Universiti Malaysia Perlis, Malaysia.}
\affiliation{Department of Physics, Cross River University of Technology, Calabar, Nigeria.}

\author{F. C. E. Lima}
\email[]{E-mail: cleiton.estevao@fisica.ufc.br}
\affiliation{Department of Physics, Universidade Federal do Cear\'{a}, Campus do Pici, Fortaleza, CE, Brazil.}
	
\author{C. A. S. Almeida}
\email[]{E-mail: carlos@fisica.ufc.br}
\affiliation{Department of Physics, Universidade Federal do Cear\'{a}, Campus do Pici, Fortaleza, CE, Brazil.}

\author{N. Ali}
\email[]{E-mail: norshamsuri@unimap.edu.my}
\affiliation{Faculty of Electronic Engineering Technology, Universiti Malaysia Perlis, Malaysia.}

\author{M. Asjad}	
\affiliation{Department of Mathematics, Khalifa University, Abu Dhabi 127788, United Arab Emirates.}

\vspace{1cm}
\begin{abstract}
\vspace{0.25cm}
We investigate the quantum information by a theoretical measurement approach of an Aharanov-Bohm (AB) ring with Yukawa interaction in curved space with disclination. It is obtained the so-called Shannon entropy, through the eigenfunctions of the system. The quantum states considered come from a Schroedinger theory with the AB field in the background of curved space. With this entropy it can explored the quantum information at the position space and reciprocal space.  Furthermore, we discussed how the magnetic field, the AB flux, and the topological defect influence the quantum states and the information entropy.
\end{abstract}

\maketitle

\section{Introduction}
Structures with the profile of quantum rings attract the attention of several researchers \cite{1,2,3,4} due to their various technological applications, e. g., nano-flash memories \cite{5,6}, photonic detectors \cite{5, 7}, and spintronics \cite{5}. Generally, this type of system classifies into two classes, i. e., the one-dimensional rings (configurations with constant radius) \cite{8, 9, 10, 11,12} and the two-dimensional rings (with variable radius) \cite{13,14,15,16}. A particular category of these structures is the Aharanov-Bohm (AB) rings. Indeed, many works study AB rings \cite{17, 18, 19}. In summary, we can understand the AB ring as structures produced by particles in a circular motion and subjected to the AB field. Actually, there are works in the literature discussing these structures with mesoscopic decoherence \cite{20}, electromagnetic resonator \cite{21}, and spin-orbit interaction \cite{22, 23}.

A physical model of great interest is a system composed of particles of spin-zero \cite{24, 25, 26}. In these systems, there is a screened Coulomb potential known as the Yukawa potential (YP) \cite{27, 28}. This potential has the profile
\begin{equation}
    V(r)=-\frac{V_1 e^{-\delta r}}{r},
\end{equation}
where the parameter $V_1$ is a coupling constant that regulates the magnitude of the effective force and the parameter $\delta$ is a parameter that makes the exponential argument dimensionless. Naturally, this potential is central and attractive. These characteristics of this potential make the interest in this topic growing \cite{29,30,31,32,33}. In his seminal work, H. Yukawa shows that this potential results in the interaction of a massive scalar field with a massive bosonic field \cite{34}. The fact is that today's theories with Yukawa's interaction have several applications \cite{35,36,37}. A stimulating application is an interaction between two nuclei. In this case, this application is interesting because two cores can experience attractive interaction due to the interaction of charged pions \cite{34}. In other words, pions are similar to two particles interacting electromagnetically through the exchange of photons.

Not far from quantum theory, information theory has been a practical tool to investigate uncertainty measurements related to quantum-mechanical systems \cite{38,39,40,41,42}. Information theory emerged with Shannon's seminal paper {\it A Mathematical Theory of Communication} in 1948 \cite{43}. Shannon sought to understand the propagation of information in a noisy channel. Consequently, he sought to explain the possible savings due to the statistical structure of the original message and due to the nature of the final destination of the message \cite{43}. Analyzing the likely interference events in the information, Shannon proposes the entropic quantity
\begin{equation}
    S=-\sum_i \rho_i\text{ln}\rho_i,
\end{equation}
where $\rho_i$ is the probability density associated with the event. Shannon's theory has also provided support for the cryptography \cite{44}, and for the noise theory \cite{45}.

We seek in this study to answer the issue: How the magnetic field, the AB flux, and the topological defect (i.e., the disclination defect) will influence a particle restricted to a Yukawa-like potential. To reach our purpose, we numerically investigate the Shannon entropy. This study is influential because it allows us to predict how the measurement uncertainties will change as the magnetic field, AB flux, or topological defect varies.

The paper is organized as follows. In Sec. II, we build the model considering a particle confined by a Yukawa interaction. Furthermore, the model was structured considering a disclination defect. In Sec. II, we exposed the analytical solutions of the quantum eigenstates. Posteriorly, in Sec. III, the numerical result of Shannon entropy is discussed. Finally, we close by announcing in Sec. IV our findings.

\section{Theory and solutions}\label{SecII}

Let us consider that the particle is confined by a Yukawa potential (YP), under the complete effect of the AB and magnetic fields. Let us assume that there exists a disclination or topological defect in this region. The disclination is described by the line element \cite{46}
\begin{equation}
    ds^2=dr^2+\alpha^2 r^2d\phi^2+dz^2
\end{equation}
where $0<\alpha<1$ is the parameter associated with the deficit of angle. The parameter $\alpha$ is related to the linear mass density $\hat{\mu}$ of the string via $\alpha=1-4\hat{\mu}$ \cite{47}. Notice that the azimuthal angle is defined in the range $0\leq \phi\leq 2\pi$ \cite{48}.

The Hamiltonian operator of a particle that is charged and confined to move in the region of YP under the collective impact of AB flux and an external magnetic fields with topological defect can be written in cylindrical coordinates. Thus, the Schr\"{o}dinger equation for this consideration is written as follows \cite{49};
\begin{equation}
    \left[\frac{1}{2\mu}\bigg(i\hbar\vec{\nabla}_{\alpha}-\frac{e}{c}\vec{A}_\alpha\bigg)^2-\frac{V_1\e^{-\delta r}}{r}\right]\psi(r,\phi)=E_{nm}\psi(r,\phi),
\end{equation}
where $E_{nm}$ denotes the energy level, $\mu$ is the effective mass of the system, the vector potential which is denoted by ``$\vec{A}$'' can be written as a superposition of two terms $\vec{A}=\vec{A}_1+\vec{A}_2$ having the azimuthal components \cite{49} and external magnetic field with $\vec{\nabla}\times\vec{A}_1=\vec{B}$, $\vec{\nabla}\times\vec{A}_2=0$, where
$\vec{B}$ is the magnetic field. $\vec{A}_1=\frac{\vec{B}\e^{-\delta r}}{\alpha(1-\e^{-\delta r})}\hat{\phi}$ and $\vec{A}_2=\frac{\Phi_{AB}}{2\pi r}\hat{\phi}$ represents the additional magnetic flux $\Phi_{AB}$ created by a solenoid with $\vec{\nabla}\cdot\vec{A}_2=0$. The Del and Laplacian operator are used as in Ref. \cite{49}.

The vector potential in full is written in a simple form as 
\begin{equation}\label{VP}
    \vec{A}_\alpha=\bigg(0,\frac{\vec{B}\e^{-\delta r}}{\alpha(1-\e^{-\delta r})}+\frac{\Phi_{AB}}{2\pi r},0\bigg).
\end{equation}

\begin{figure*}
    \centering
    \includegraphics[width=0.4\textwidth]{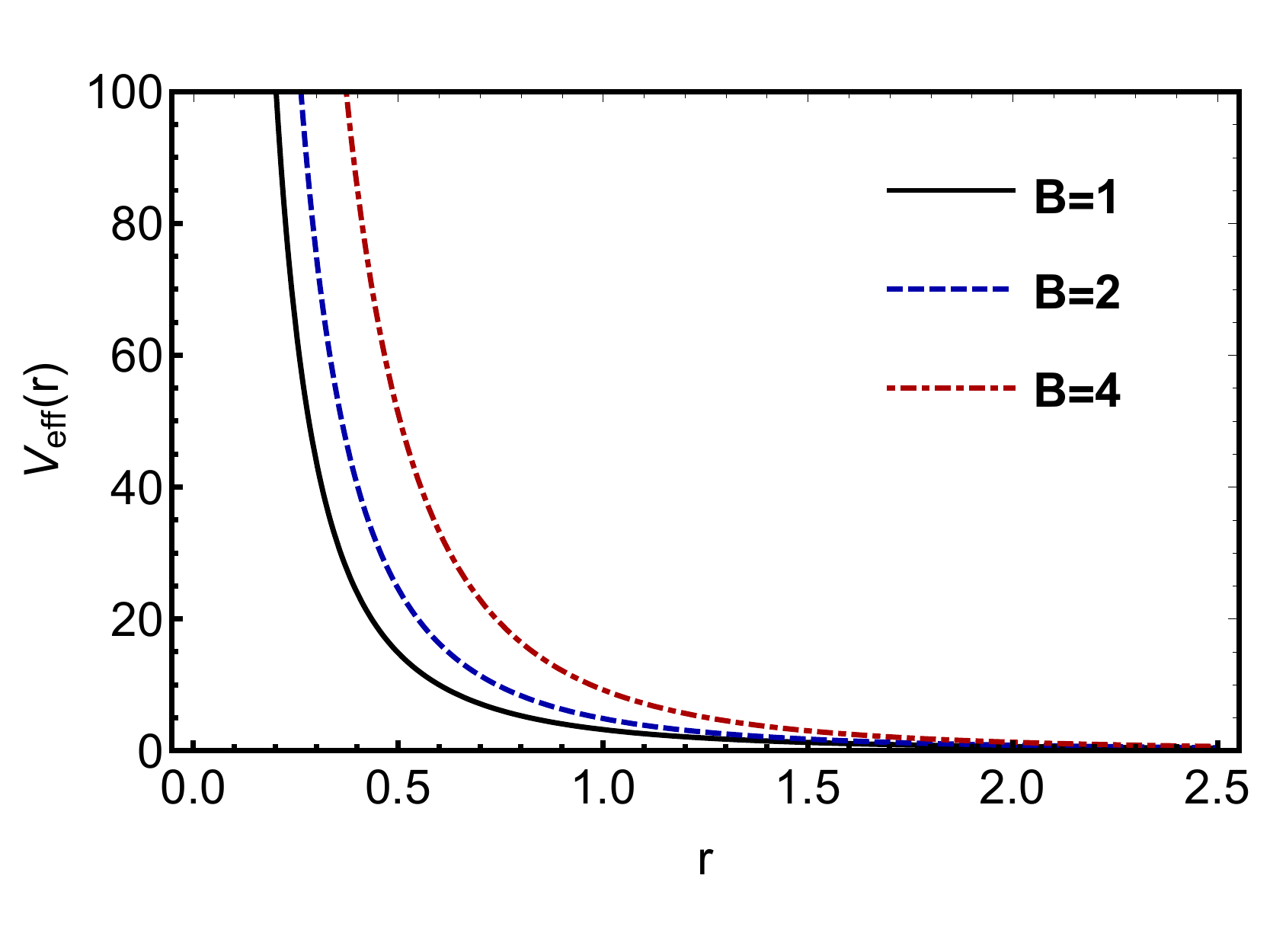}
    \includegraphics[width=0.4\textwidth]{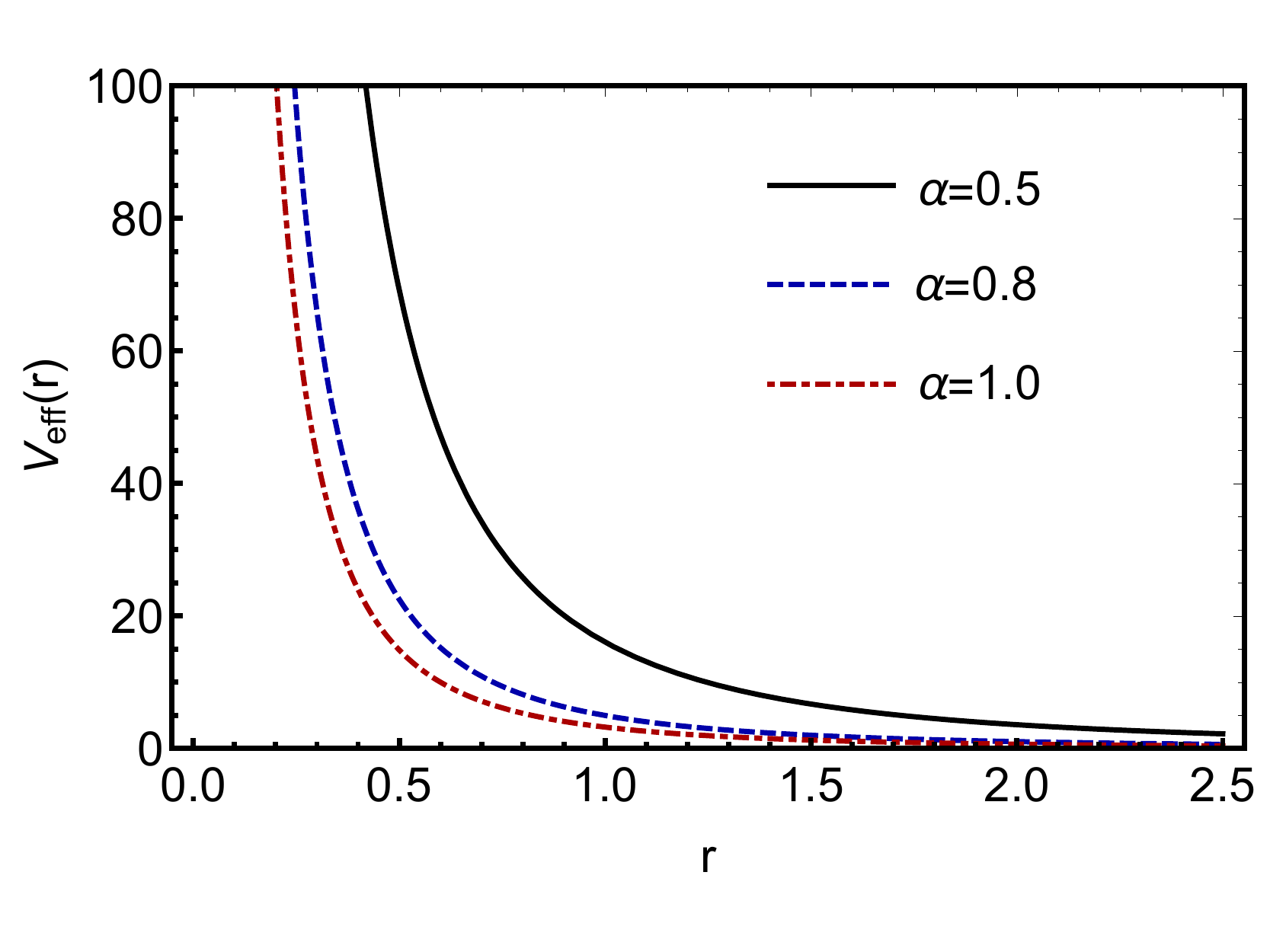}\\
    \vspace{-0.5cm}
    \begin{center}
        (a) \hspace{8cm} (b)
    \end{center}
     \includegraphics[width=0.4\textwidth]{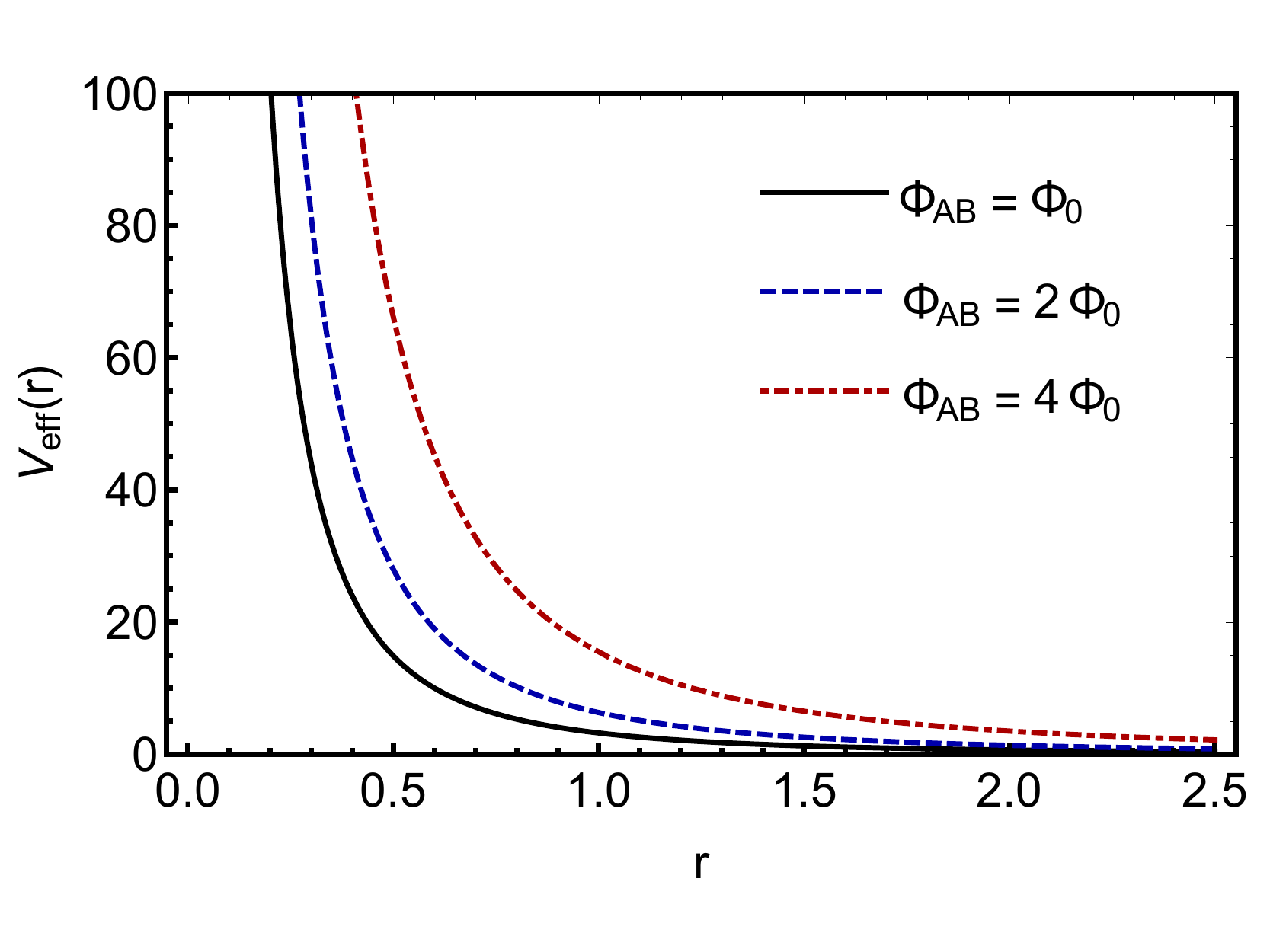}\\ \vspace{-0.5cm} \begin{center}
         (c)
     \end{center}
     \vspace{-0.7cm}
\caption{(a) Effective potential when the magnetic field changes. (b) Effective potential when the parameter $\alpha$ (topological defect) varies. (c) Effective potential when flux AB varies.}
    \label{fig1}
\end{figure*}

Let us take a wave function in the cylindrical coordinates as
\begin{equation}
    \psi(r,\phi)=\frac{1}{\sqrt{2\pi r }}\e^{im\phi}R_{nm}(r),
\end{equation}
where $m$ denotes the magnetic quantum number. Inserting this wave function and the vector potential into Eq. (\ref{VP}) we arrive at the following radial second-order differential equation:
\begin{equation}\label{REquation}
    R''_{nm}(r)+\frac{2\mu}{\hbar^2}[E_{nm}-V_{\text{eff}}(r)]R_{nm}(r)=0,
\end{equation}
where $V_{\text{eff}}(r)$ is the effetive potential defined as
\begin{align} \label{VEff}
 V_{\text{eff}}(r)=&-\frac{V_1\e^{-\delta r}}{r}+\hbar\omega_c\left(\frac{m}{\alpha^2}+\frac{\xi}{\alpha}\right)\frac{\e^{-\delta r}}{(1-\e^{-\delta r})r}+\left(\frac{\mu\omega_c^2}{2}\right)\frac{\e^{-2\delta r}}{(1-\e^{-\delta r})^2}+\frac{\hbar^2}{2\mu}\left[\frac{(\frac{m}{\alpha^2}+\xi)^2-\frac{1}{4}}{r^2}\right],
\end{align}
where $\xi=\frac{\Phi_{AB}}{\Phi_0}$ is an integer with the flux quantum $\Phi_0=\frac{hc}{e}$ and $\omega_c=\frac{e\vec{B}}{\mu c}$ denotes the cyclotron frequency. Fig. \ref{fig1} is displayed the behavior of the effective potential when the magnetic field, the topological defect, and the AB flux undergo changes.

Eq. (\ref{REquation}) is a complicated differential equation that cannot be solvable easily due to the presence of centrifugal term. Therefore, we employ the Greene and Aldrich approximation scheme \cite{50} to bypass the centrifugal term. This approximation is given
\begin{equation}\label{Approx}
    \frac{1}{r^2}=\frac{\delta^2}{(1-\e^{-\delta r})^2}.
\end{equation}
We point out here that this approximation is only valid for small values of the screening parameter $\delta$.

Inserting Eqs. (\ref{Approx}) into Eq. (\ref{VEff}) and introducing a new variable $s=\e^{-\delta r}$ allows us to obtain
\begin{eqnarray}
    \frac{d^2R_{nm}}{ds^2}+\frac{1}{s}\frac{dR_{nm}(s)}{ds}+\frac{1}{s^2(1-s^2)^2}[-(\varepsilon_{nm}+\beta_0+\beta_2)s^2+(2\varepsilon_{nm}+\beta_0-\beta_1)s-(\varepsilon_{nm}+\eta)]R_{nm}(s)=0.&
    \label{EqR}
\end{eqnarray}

For Mathematical simplicity, let’s introduce the following dimensionless notations;
\begin{eqnarray}\label{rel1}
    \varepsilon_{nm}=-\frac{2\mu E_{nm}}{\hbar^2\delta^2}, \, \, \, \, \, \,  \beta_0=\frac{2\mu V_1}{\hbar^2 \delta}, \, \, \, \, \, \, \beta_1=\frac{2\mu\omega_c}{\hbar\delta}\bigg(\frac{m}{\alpha^2}+\frac{\xi}{\alpha}\bigg), \, \, \, \, \, \, \beta_2=\frac{\mu^2\omega_{c}^{2}}{\hbar^2\delta^2}, \, \, \, \, \, \text{and} \, \, \, \, \, \, \eta=\bigg(\frac{m}{\alpha}+\xi\bigg)^2-\frac{1}{4}.
\end{eqnarray}
In order to solve eq. (\ref{EqR}), we have transform the differential equation (8) into a form solvable by any of the existing standard mathematical technique. Hence, we take the radial wave function of the form
\begin{equation}\label{Trans}
R_{nm}(s)=s^{\lambda}(1-s)^{\nu}f_{nm}(s),
\end{equation}
where $\lambda=\sqrt{\varepsilon_{nm}+\eta}$, and  $\nu=\frac{1}{2}+\sqrt{\frac{1}{4}+\beta_1+\beta_2+\eta}$.
On substitution of Eq. (\ref{Trans}) into Eq. (\ref{EqR}), we obtain the following hypergeometric equation:
\begin{eqnarray}
    s(1-s)f''_{nm}(s)+[(2\lambda+1)-(2\lambda+2\nu+1)s]f'_{nm}(s)-[(\lambda+\nu)^2-(\sqrt{\varepsilon_{nm}+\beta_0+\beta_2})^2]f_{nm}(s)=0.&
\end{eqnarray}

By considering the finiteness of the solutions, the quantum condition is given by
\begin{equation}
   (\lambda+\nu)-(\sqrt{\varepsilon_{nm}+\beta_0+\beta_2})=-n
\end{equation}
which in turn transforms into the energy eigenvalue equation follows;
\begin{eqnarray} \label{e1}
  \varepsilon_{nm}=&-\eta+\frac{1}{4}\left[\frac{\beta_0+\beta_2-\eta-\bigg(n+\frac{1}{2}+\sqrt{\frac{1}{4}+\beta_1+\beta_2+\eta}\bigg)^2}{n+\frac{1}{2}+\sqrt{\frac{1}{4}+\beta_1+\beta_2+\eta}}\right]^2.
\end{eqnarray}
Substituting Eq. (\ref{rel1}) into Eq. (\ref{e1}) and carrying some simple manipulative algebra, we arrive at the energy eigenvalue equation of the Yukawa potential in the presence of magnetic and AB fields with topological defect in the form
\begin{align}\nonumber
    E_{nm}=&\frac{-(\hbar^2\delta^2/8\mu)}{\bigg[n+\frac{1}{2}+\sqrt{\frac{\mu^2\omega_c^2}{\hbar^2\delta^2}+\frac{2\mu\omega_c}{\hbar\delta}\bigg(\frac{m}{\alpha^2}+\frac{\xi}{\alpha}\bigg)+\bigg(\frac{m}{\alpha}+\xi\bigg)^2}\bigg]}\bigg\{\frac{2\mu V_1}{\hbar^2\delta}+\frac{\mu^2\omega_c^2}{\hbar^2\delta^2}-\eta-\bigg[n+\frac{1}{2}+\\
    +&\sqrt{\frac{\mu^2\omega_c^2}{\hbar^2\delta^2}+\frac{2\mu\omega_c}{\hbar\delta}\bigg(\frac{m}{\alpha^2}+\frac{\xi}{\alpha}\bigg)+\bigg(\frac{m}{\alpha}+\xi\bigg)^2}\bigg]^2\bigg\}+\frac{\hbar^2\delta^2\eta}{2\mu}.
\end{align}

Furthermore, the wave eigenfunction is 
\begin{eqnarray}\label{wave}
    \psi_{nm}(r,\phi)=&\frac{1}{\sqrt{2\pi r}}\e^{im\phi} \, _{1}F_{1}(a, b, c;\e^{-\delta r}) \e^{-\delta r\sqrt{\varepsilon_{nm}+\eta}}
    (1-\e^{-\delta r})^{\frac{1}{2}+\sqrt{\frac{1}{4}+\beta_1+\beta_2+\eta}},  
\end{eqnarray}
where $a=\lambda+\nu+\sqrt{\varepsilon_{nm}+\beta_0+\beta_2}$, $b=\lambda+\nu-\sqrt{\varepsilon_{nm}+\beta_0+\beta_2}$, and $c=2\lambda+1$.

\begin{figure*}
\centering
\includegraphics[width=0.4\textwidth]{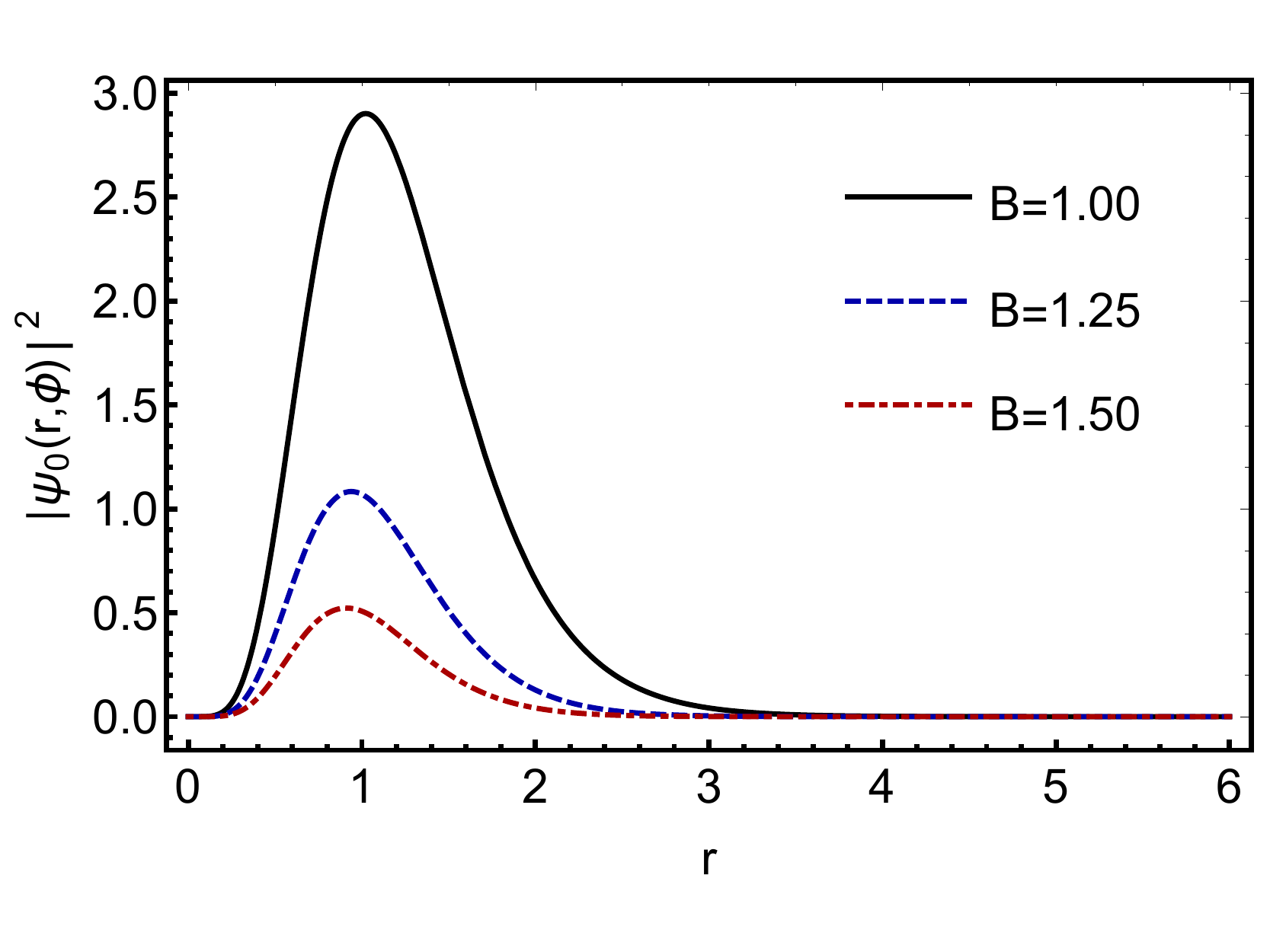}
\includegraphics[width=0.4\textwidth]{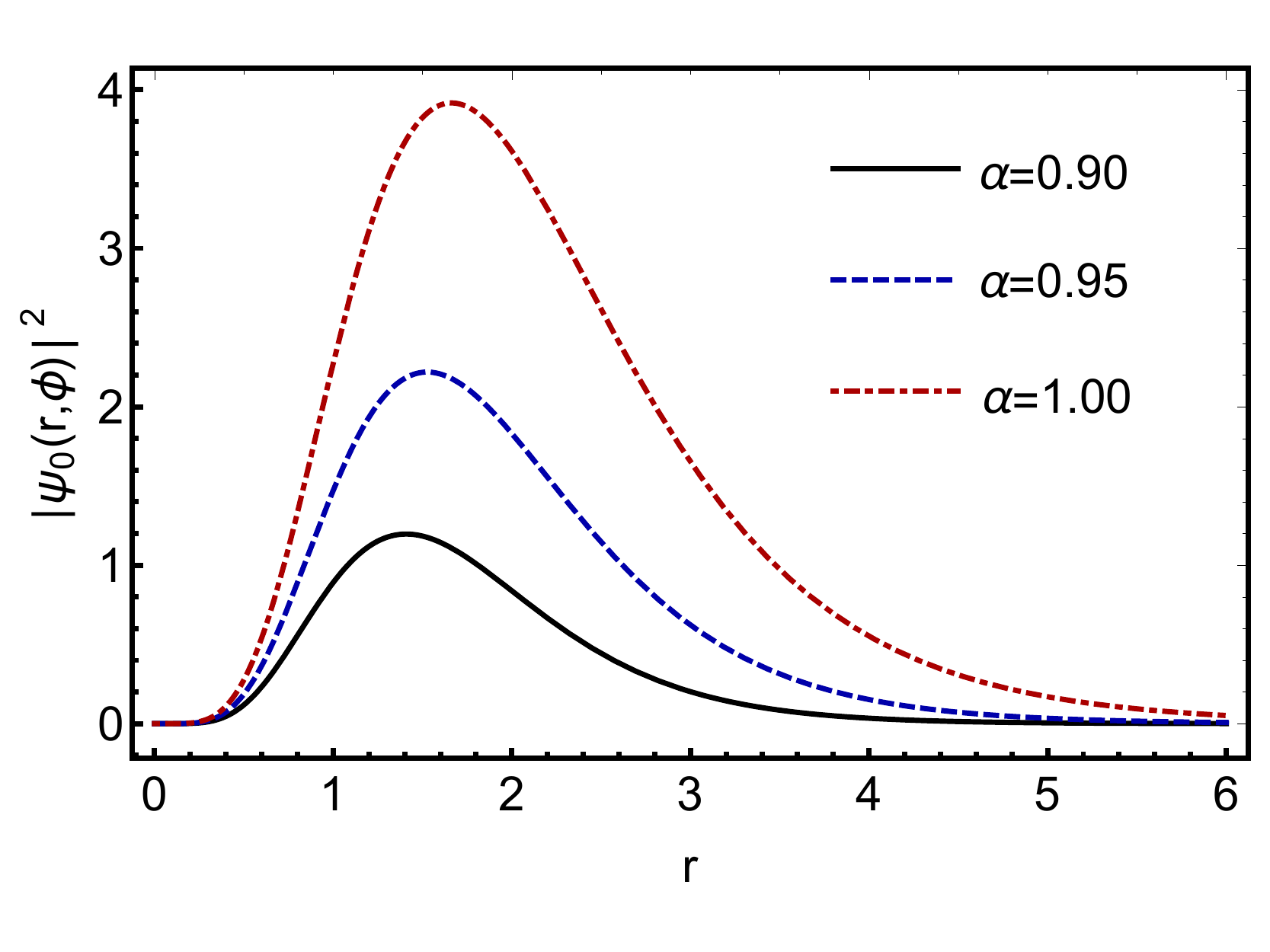}\\
\vspace{-0.5cm}
\begin{center}
    (a) \hspace{8cm} (b)
\end{center}
\vspace{-0.5cm}
\includegraphics[width=0.4\textwidth]{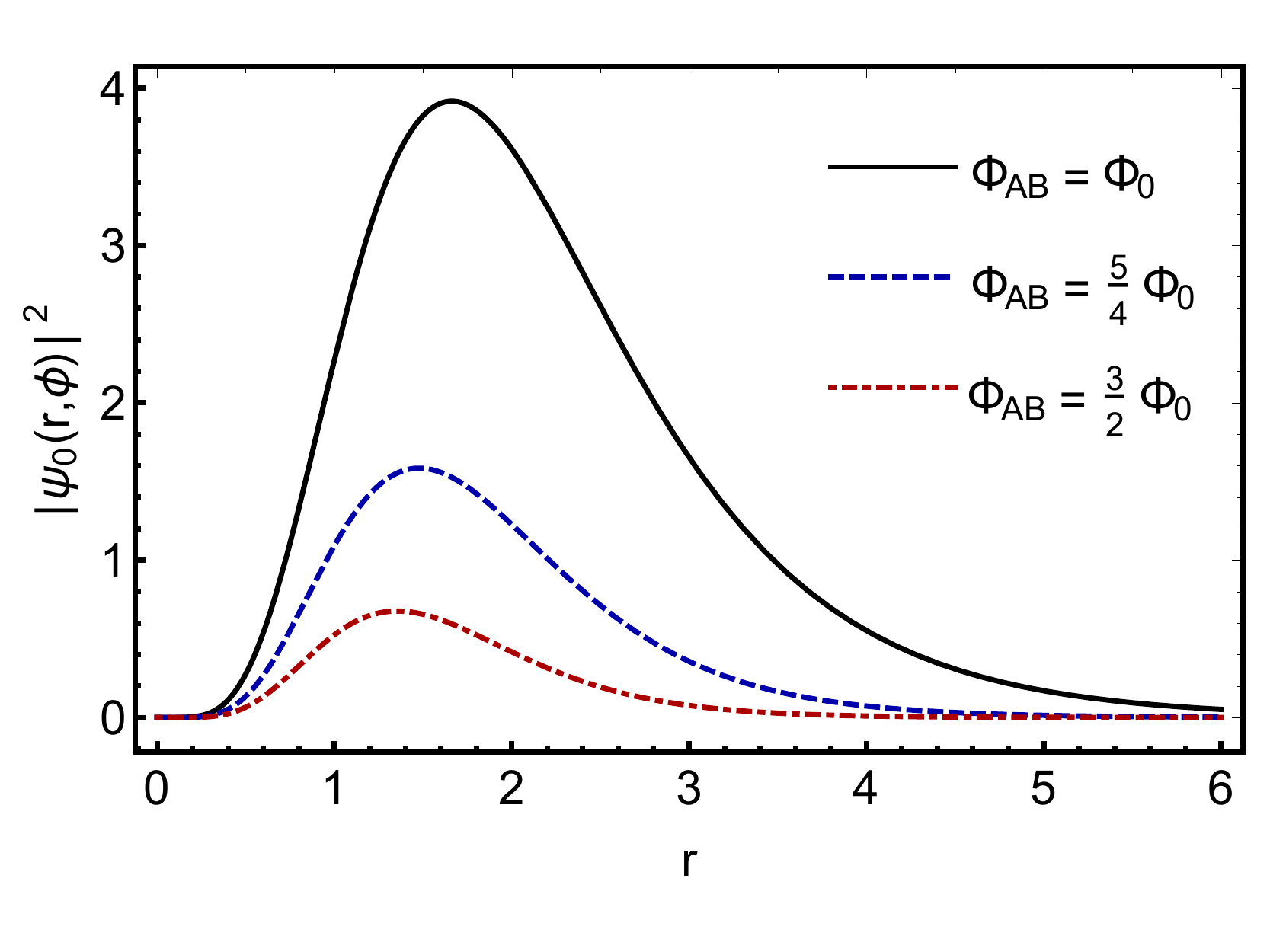}\\
\begin{center} (c) \end{center}
\vspace{-0.7cm}
\caption{ Probability density $|\psi(r)|^2$ in position space $r$ when magnetic field changes (a), the parameter $\alpha$ (topological defect) varies (b) and AB flux varies (c). }
\label{fig2}
\end{figure*}

In Fig. \ref{fig2} are displayed the plots of the probability density of the particle confined to the Yukawa interaction (Eq. (\ref{VEff})) in the presence of a disclination defect. Note that the wave function profile changes when the magnetic field, the AB flux, and the topological defect change in the theory. Furthermore, when the declination defect (i.e., when $\alpha>1$) is changed, the probability density of the wave function translates in its profile so that the highest probability is close to the atomic core. Here, it is interesting to mention that the excited eigenstates of the model will have a profile similar to that shown in Fig. \ref{fig2}. However, a change in the height of the solitonic wave function can be verified in excited states.
\section{Shannon's entropy}
Quantum information entropy has helped to understand the physics of several systems \cite{51, 52, 53}. Indeed, one of the quantum entropies used to study the information of quantum systems is the Shannon entropy. Shannon's entropy emerged within the scope of information theory seeking to describe the best way to propagate information between a source and a receiver \cite{43}. Along with the development of the Shannon entropy concept in physics, studies on the thermodynamics of an ensemble of particles led to mathematical expressions with a similar profile. This similarity of Shannon information and Boltzmann entropy allowed Shannon information to be called Shannon entropy \cite{54}. Some conceptual applications of Shannon's entropy help us understand the information and uncertainty measurement of quantum systems, e. g., the Shannon entropy gives us the uncertainty of non-Hermitian particle systems \cite{55}. Furthermore, Shannon formalism allowed the study of fermionic particles \cite{56}, problems with effective mass distribution \cite{57, 58}, and mechanical-quantum models with double-well potential \cite{59}.

An interpretation of the Shannon entropy of a quantum-mechanical system tells us the measure of uncertainties of a probability distribution associated with a source of information \cite{60, 61}. The Born interpretation of the quantum mechanics \cite{62} takes us to the statistical perception of the stationary quantum system, i. e., 
\begin{align}\label{rho}
    \rho(\vec{r})\, dr=\vert\psi(\vec{r},t)\vert^2\, dr\equiv\vert \psi(\vec{r})\vert^2\, dr.
\end{align}
In this case, $\rho(\vec{r})$ is the probability of finding the particle in the state $\psi(\vec{r},\,t)$ between $\vec{r}$ and $\vec {r}+d\vec{r}$ \cite{62}. Furthermore, $\vert\psi(\vec{r},\, t)\vert^2$ is the probability density of the quantum-mechanical system.

Let us now study Shannon's entropy in the context of quantum mechanics. Remembering that the probability density has the form of Eq. (\ref{rho}), we define Shannon's entropy as
\begin{align}
    S=-\sum_{i}\rho_i\,\text{ln}\rho_i,
\end{align}
so that for a probability density $\rho_i$ of a continuous system in position space, Shannon entropy takes the form
\begin{align}\label{Sx}
    S_r=-\int_{-\infty}^{\infty}\,\vert\psi(r)\vert^{2}\,\text{ln}(\vert\psi(r)\vert^{2})\, dr.
\end{align}

On the other hand, Shannon's entropy in reciprocal space (or momentum space) is
\begin{align}\label{Sp}
    S_k=-\int_{-\infty}^{\infty}\,\vert\psi(k)\vert^{2}\,\text{ln}(\vert\psi(k)\vert^{2})\, dk.
\end{align}
Perceive that the expressions (\ref{Sx}) and (\ref{Sp}) have only two independent variables. Our model is two-dimensional, i. e., a model of (2+1)D with spatial coordinates $r$ and $\phi$. However, $\phi$ is a cyclic coordinate. In this way, only the $r$ coordinate contributes to Shannon information.

The wave function in reciprocal space $\psi(k)$ is given by Fourier transform, namely,  
\begin{align}\label{FT}
    \psi(k)=\frac{1}{\sqrt{2\pi}}\int_{-\infty}^{\infty}\,\psi(r)\,\text{e}^{-irk}\,dr.
\end{align}
The entropic quantities of Eqs. (\ref{Sx}) and (\ref{Sp}) play a role analogous to the Heisenberg uncertainty measures \cite{55, 56}. An entropic uncertainty relation that relates to the entropic uncertainties was obtained by Beckner \cite{61}, Bialynicki-Birula and Mycielski (BBM) \cite{63}. The BBM relation of uncertainties is 
\begin{align}
    S_r+S_k\geq \text{D}(1+\text{ln}\pi),
\end{align}
where D is the dimension of effective spatial coordinates, i. e., the number of spatial coordinates that contribute to the information propagation. In our case, the results must respect the relation
\begin{align}
    S_r+S_k\geq 2.14473.
\end{align}
To investigate the quantum information of the model presented in Sec. \ref{SecII}, we consider the wave function (\ref{wave}) and numerically normalize it through the expression
\begin{align}
    \int_{-\infty}^{\infty}\, \vert\psi_{nm}(r,\phi)\vert^2\, dr=1.
\end{align}
After, we use Eq. (\ref{Sx}) to calculate Shannon entropy in position space. Using the Fourier transform (\ref{FT}), it comes to eigenfunctions at the reciprocal space. With the eigenfunctions normalized in the momentum space, we use Eq. (\ref{Sp}) to obtain Shannon's entropy in this space. The numerical results of Shannon's entropy for the first energy levels are in tables \ref{tab1} and \ref{tab2}. We presented the physical discussion of the results found in the next section.
\begin{table}
\caption{Numerical result of Shannon's entropy for several values of the magnetic field and flux AB.}
\resizebox{8cm}{!}{%
\begin{tabular}{|c|c|c|c|c|c|c|} \hline
$n$ & $m$ & $B$ & $\Phi_{AB}$ & $S_r$ & $S_k$ & $S_r+S_k$ \\ \hline
 \multirow{6}{*}{0} & \multirow{6}{*}{0} & 1 & 1 & 1.32078 & 2.91721 & 4.23799\\ \cline{3-7} 
  &  & 2 & 1  & 0.69776 & 3.53949 & 4.23725\\  \cline{3-7}
  &  & 4 & 1 & 0.20082 & 4.37062 & 4.57144\\ \cline{3-7}
  &  & 1 & 2 & 0.68816 & 3.24081 & 3.92897\\  \cline{3-7}
  &  & 1 & 4 & 0.20081 & 5.93350 & 6.13431\\ \hline
\multirow{6}{*}{1} & \multirow{6}{*}{0} & 1 & 1 & 4.41786 & 7.14836 & 11.56622\\  \cline{3-7}
  &  & 2 & 1 & 0.53510 & 10.14113 & 10.67623 \\  \cline{3-7} 
  &  & 4 & 1 & 0.41849 & 10.44393 & 10.86242 \\  \cline{3-7} 
  &  & 1 & 2 & 0.83093 & 4.83668  & 5.66761 \\  \cline{3-7} 
  &  & 1 & 4 & 0.19793 & 6.30831 & 6.50624 \\  \hline
  
\multirow{6}{*}{1} & \multirow{6}{*}{1} & 1 & 1 & 6.52497 & 8.45892 & 14.98389\\  \cline{3-7}
  &  & 2 & 1 & 0.36401 & 11.48506 & 11.84907 \\ \cline{3-7} 
  &  & 4 & 1 & 0.31416 & 11.78786 & 12.10202 \\  \cline{3-7} 
  &  & 1 & 2 & 0.35912 & 6.29333 & 6.65245 \\  \cline{3-7} 
  &  & 1 & 4 & 0.03149 & 6.36141 & 6.39290 \\  \hline
\end{tabular}} \label{tab1}
\end{table}
\begin{table}
\caption{Numerical result of Shannon's entropy when the disclination varies.}
\resizebox{8cm}{!}{%
\begin{tabular}{|c|c|c|c|c|c|} \hline
$n$ & $m$ & $\alpha$ & $S_r$ & $S_k$ & $S_r+S_k$ \\ \hline
\multirow{3}{*}{0} & \multirow{3}{*}{0} & 0.1 & -1.43473 & 5.30857 & 3.87384 \\ \cline{3-6}
 &  & 0.2 & -0.73813 & 3.14145 & 2.40332 \\ \cline{3-6}
 &  & 0.4 & 1.24625 & 1.96739 & 3.21364 \\ \hline
 \multirow{3}{*}{1} & \multirow{3}{*}{0} & 0.1 & -1.52895 & 8.04484 & 6.51589 \\ \cline{3-6}
 &  & 0.2 & -0.91699 & 6.30221 & 5.38522 \\ \cline{3-6}
 &  & 0.4 & -0.05982 & 4.30012 & 4.24030 \\ \hline
 \multirow{3}{*}{1} & \multirow{3}{*}{1} & 0.1 & -1.60572 & 9.35928 & 7.75356 \\ \cline{3-6}
 &  & 0.2 & -0.96676 & 9.14016 & 8.17340 \\ \cline{3-6}
 &  & 0.4 & -0.31977 & 5.52721' & 5.20744 \\ \hline
\end{tabular}} \label{tab2}
\end{table}

\section{Final remarks}
Throughout the paper, we studied the influences of the external magnetic field, the AB flux, and the disclination defect on the AB quantum ring. The wave function that describes the system is the confluent hypergeometric function. Furthermore, the complete solution set, i. e., the wave eigenfunctions, reproduces a null probability at the spatial infinity. This condition leads us to a discretized energy spectrum indicating the existence of bound states.

We measure the Shannon entropy to analyze the quantum information. It was possible to notice that the disclination, the external magnetic field, and AB flux directly influence the quantum information of the system. Considering the numerical results displayed in tables \ref{tab1} and \ref{tab2}, it is notorious that the informational content decreases in the position space when the contribution of the AB flux to the magnetic field increases. This is because the contribution of the magnetic field and the AB field, when increasing their intensity, make the quantum rings more localized, so the informational content decreases. Indeed, this indicates a decrease in the uncertainties related to the measurements of the position of the AB ring. In counterpoint, it is prominent that if the disclination increases, quantum information increases (in position space). Therefore, the increases in information (or uncertainties) in the position space is grow up. Thereby, this is a consequence of the type of defect. In the absence of disclination (i. e., $\alpha=1$), position uncertainty is greater. However, occurring the rotational symmetry breaking, the position measurement uncertainties begin to decreases. In this case, this is because the ring became thinner and thinner.

On the other hand, quantum information (and consequently, measurement uncertainty) increases at the momentum space. The measurement uncertainties increase as the magnetic field and AB flux increase. However, when topological defect increase, the information decrease in reciprocal space. The increment (or reduction) of the information at the momentum space is a consequence of the Heisenberg uncertainty principle (HUP). Moreover, in the communication theory, the BBM relation plays the role of the HUP. Here is essential to mention that BBM relation is valid in our model.

A direct perspective of this work is the study of the quantum information measurement of a quantum ring in the presence of other topological defects. Another possibility is to study the relativistic version of this theory. We hope to produce these studies in the future.

\section*{Acknowledgements} 
This research has been carried out under LRGS Grant 9012-00009 Fault-tolerant Photonic Quantum States for Quantum Key Distribution provided by Ministry of Higher Education of Malaysia (MOHE) and by Khalifa University through project no. 8474000358 (FSU-2021-018). F. C. E. Lima is grateful the Coordena\c{c}\~{a}o de Aperfei\c{c}oamento do Pessoal de N\'{i}vel Superior (CAPES), for the financial support. C. A. S. Almeida thanks the Conselho Nacional de Desenvolvimento Cient\'{\i}fico e Tecnol\'{o}gico (CNPq), for the financial support in the project n$\textsuperscript{\underline{\scriptsize o}}$ 309553/2021-0.
\bibliographystyle{unsrt}

\end{document}